\documentclass[onecolumn,tightenlines,showpacs,nofootinbib,preprintnumbers,
amsmath,amssymb]{revtex4}

\usepackage[usenames,dvipsnames]{color}
\usepackage{caption2}
\usepackage{dcolumn}
\usepackage{graphicx}
\usepackage{subfigure}
\usepackage{epstopdf}
\usepackage{bm}
\usepackage{lscape}

\usepackage{setspace}
\begin{document}
\begin{spacing}{1.5}

\title{Isospin symmetry breaking effects for the mode ${B}\rightarrow \pi\pi(K)$ in Perturbative QCD}

\author{Gang L\"{u}$^{1}$\footnote{Email: ganglv66@sina.com},
Liang-Chen Liu $^{1}$\footnote{Email: llc163mail@163.com},
 Qin-Qin Zhi}

\affiliation{\small $^{1}$College of Science, Henan University of Technology, Zhengzhou 450001, China\\
}

\begin{abstract}
We calculate the direct $CP$ violation for the decay process of
${B}^{0}\rightarrow \pi^{0}\pi^{0}$, ${B}^{+}\rightarrow \pi^{0}\pi^{+}$, ${B}^{0}\rightarrow K^{0}\pi^{0}$ and ${B}^{+}\rightarrow K^{+}\pi^{0}$ via isospin symmetry breaking effects from the $\pi^{0}-\eta-\eta'$ mixing mechanism in perturbative QCD  approach.
Isospin symmetry breaking originates from the electroweak interaction and the u-d quark mass difference through the strong interaction
which are known to be tiny. However, we find that isospin symmetry breaking at the leading order changes
the $CP$ violation from the new strong phases. Our calculation results for the $CP$ violation are within or including the range of experimental results.
We also compare our results with those from the QCD factorization and the perturbative QCD schemes without regard to isospin symmetry breaking effects.

\end{abstract}

\maketitle

\section{\label{intro}Introduction}
The theoretical studies and experiment of the two body charmless hadronic B meson decay play an important part in searching new physics signals beyond the Standard Model (SM). In the SM, the non-leptonic decays of ${B}\rightarrow \pi\pi$ and ${B}\rightarrow \pi K$ provide valuable information on the inner angles of the unitarity triangle of Cabibbo-Kobayashi-Maskawa (CKM)\cite{cab, kob} matrix, and have been widely studied \cite{SB}. Due to the quark flavor mixing, the CKM matrix  provides us the weak phases. Associated with a strong stage, the weak phase is the source of CP violation. The strong phase comes from the dynamics of QCD and
the other mechanism.

Since the interference of tree and penguin amplitudes is expected to produce $CP$ violation, the charmless  ${B}\rightarrow \pi\pi$ and ${B}\rightarrow \pi K$ modes are very representative in the two-body B meson decays, and hence have been studied most extensively. Due to the interaction between the theory and data of B factories, it is possible to obtain valuable new insights into the physics of these modes,
and also raises a question about the field of electroweak penguin sector,  which can be contributed according to the new physics to modify. \cite{R.}. There are some potential contradiction between the current new B-factory data for ${B}\rightarrow \pi\pi$ and ${B}\rightarrow \pi K$  decays and the predictions. For example, new experimental data for ${B}^{0}\rightarrow \pi^{0}\pi^{0}$ and  ${B}^{0}\rightarrow K^{0}\pi^{0}$  decay rates are significantly larger than the theoretical predictions \cite{Li}. In addition, the prediction of direct CP asymmetry in these modes is also inconsistent with the data, even if the signs of some processes are reversed \cite{MB,MB1}.
There is conspicuous difference for the $CP$ violation between the process $B^{\pm}\rightarrow \pi^{0}K^{\pm}$ and $B^{0}\rightarrow \pi^{\mp}K^{\pm}$.
For the scheme of QCD factorization, the perturbative  and non-perturbative contribution are suppressed by $\mathcal{O}(\alpha_{s})$ and $\mathcal{O}(\Lambda_{QCD})$,
respectively. Therefore, because there are very few violations of $CP$, it is difficult to explain the difference.
The Glauber-gluon effects are discussed for the $CP$ violation of process ${B}\rightarrow \pi K$
from the Glauber phase factors which lead to the rotation of the tree amplitude by a strong phase
in the framework of perturbative QCD \cite{liuxin2016}.
New physics contribution is the new sources of $CP$ violation through electroweak penguins for the decay process of ${B}\rightarrow \pi K$
\cite{Buras2004NB,Barger2009,Beaudy2018}.

Isospin is an approximate symmetry in the Standard model which can be broken either by electroweak effects or by the strong interaction through the u and d quark mass difference.
Both mechanisms of isospin violation lead to $\triangle I=\frac{3}{2}$ penguin contributions.
The $\pi^{0}-\eta-\eta'$ mixing arises from the latter \cite{H.} and permits an $I=1$ amplitude \cite{AJ}. These latter contributions convert the triangle relations between the amplitudes to quadrilaterals. The effect of electroweak penguin diagrams has been studied earlier in the literature and is estimated to be small \cite{NG,MG,RF}.
The quark flavor basis leads to the mixing of the pseudoscalar mesons. Isospin symmetry breaking from the QCD dynamics are
introduced by the mixing of $\pi^{0}-\eta-\eta'$, which can be applied to the phenomenology by the quark flavor mixing scheme.
The $\pi^{0}-\eta-\eta'$ mixing from u and d quark mass difference can present the
strong phase, which may affect the value of $CP$
violation accordingly. It is interesting to study how it could affect the $CP$ violation for the decay processes of ${B}\rightarrow \pi\pi$, ${B}\rightarrow \pi K $.

The remainder of this paper is organized as follows. In Sec.
\ref{sec:hamckm} we present the form of the effective Hamiltonian.
In Sec. \ref{sec:cpv1} we give the calculating formalism of $CP$ violation from isospin symmetry breaking effects.
Input parameters are presented in Sec.\ref{input}.
We present the numerical results in Sec.\ref{num}.
Summary and discussion are included in
Sec. \ref{sum}. The related functions defined in the text are given in the Appendix.

\section{\label{sec:hamckm}The effective hamiltonian}

From the operator product expansion, the effective weak Hamiltonian can
be expressed as \cite{buras}
\begin{eqnarray}
{\cal H}_{\Delta B=1} = {G_F\over \sqrt{2}}[
V^*_{ub}V_{uq}(C_1 O^u_1 + C_2 O^u_2)    \nonumber   \\
 -  V^*_{tb} V_{tq}\sum^{10}_{i=3} C_i O_i] + H.C.,\;
\label{2a}
\vspace{2mm}
\end{eqnarray}
where $G_F$ is Fermi constant, $C_i$ (i=1,...,10) represent the Wilson coefficients, $V_{ub}$,
$V_{uq}$, $V_{tb}$ and $V_{tq}(q=s,d)$ are the CKM matrix elements. The
operators $O_i$ have the following forms:
\begin{eqnarray}
\begin{split}
O^{u}_1&= \bar q_\alpha \gamma_\mu(1-\gamma_5)u_\beta\bar
u_\beta\gamma^\mu(1-\gamma_5)b_\alpha,  \\
O^{u}_2&= \bar q \gamma_\mu(1-\gamma_5)u\bar
u\gamma^\mu(1-\gamma_5)b,  \\
O_3&= \bar q \gamma_\mu(1-\gamma_5)b \sum_{q'}
\bar q' \gamma^\mu(1-\gamma_5) q',  \\
O_4 &= \bar q_\alpha \gamma_\mu(1-\gamma_5)b_\beta \sum_{q'}
\bar q'_\beta \gamma^\mu(1-\gamma_5) q'_\alpha,  \\
O_5&= \bar q \gamma_\mu(1-\gamma_5)b \sum_{q'} \bar q'
\gamma^\mu(1+\gamma_5)q',  \\
\label{2b}
\end{split}
\end{eqnarray}
\begin{eqnarray*}
\begin{split}
O_6& = \bar q_\alpha \gamma_\mu(1-\gamma_5)b_\beta \sum_{q'}
\bar q'_\beta \gamma^\mu(1+\gamma_5) q'_\alpha,  \\
O_7&= \frac{3}{2}\bar q \gamma_\mu(1-\gamma_5)b \sum_{q'}
e_{q'}\bar q' \gamma^\mu(1+\gamma_5) q',  \\
O_8 &= \frac{3}{2} \bar q_\alpha \gamma_\mu(1-\gamma_5)b_\beta \sum_{q'}
e_{q'}\bar q'_\beta \gamma^\mu(1+\gamma_5) q'_\alpha,  \\
O_9&= \frac{3}{2}\bar q \gamma_\mu(1-\gamma_5)b \sum_{q'} e_{q'}\bar q'
\gamma^\mu(1-\gamma_5)q',  \\
O_{10}& = \frac{3}{2}\bar q_\alpha \gamma_\mu(1-\gamma_5)b_\beta \sum_{q'}
e_{q'}\bar q'_\beta \gamma^\mu(1-\gamma_5) q'_\alpha.
\end{split}
\end{eqnarray*}
where $\alpha$ and $\beta$ are color indices, and $q^\prime=u, d$
or $s$ quarks. In Eq.(\ref{2b}) $O_1^u$ and $O_2^u$ are tree
operators, $O_3$--$O_6$ are QCD penguin operators and $O_7$--$O_{10}$ are
the operators associated with electroweak penguin diagrams.

We can obtain numerical values of $C_i$. When $C_i(m_b)$ \cite{LKM},
\begin{eqnarray}
\begin{split}
C_1 &=-0.2703, \;\; \;C_2=1.1188,  \\
C_3 &= 0.0126,\;\;\;C_4 = -0.0270,  \\
C_5 &= 0.0085,\;\;\;C_6 = -0.0326,  \\
C_7 &= 0.0011,\;\;\;C_8 = 0.0004,  \\
C_9&= -0.0090,\;\;\;C_{10} = 0.0022.
\label{2k}
\end{split}
\end{eqnarray}

One can obtain numerical values of $a_i$. The combinations $a_i$ of Wilson coefficients are defined as \cite{AG, YH}
\begin{eqnarray}
\begin{split}
a_1 &= C_2+C_1/ 3, \;\; \;a_2 = C_1+C_2/ 3,  \\
a_3 &= C_3+C_4/ 3,\;\;\;a_4 = C_4+C_3/ 3,  \\
a_5 &= C_5+C_6/ 3,\;\;\;a_6 = C_6+C_5/ 3,  \\
a_7 &= C_7+C_8/ 3,\;\;\;a_8 = C_8+C_7/ 3,  \\
a_9&= C_9+C_{10}/ 3,\;\;\;a_{10} = C_{10}+C_9/ 3.  \\
\label{2k}
\end{split}
\end{eqnarray}

\section{\label{sec:cpv1}$CP$ violation from Isospin symmetry breaking effects}
\subsection{\label{subsec:form}Formalism}

By containing or not containing strange quark, we can denote isospin vector triplet $\pi_{3}$, isospin scalar $\eta_{n}$ and isospin scalar $\eta_{s}$.
One can translate the $SU(3)$ singlet $\eta_{0}$ and octet $\eta_{8}$ into isospin scalar $\eta_{n}$ and isospin scalar $\eta_{s}$ by
the relations
$\eta_{n}=\frac{\sqrt{2}\eta_{0}+\eta_{8}}{\sqrt{3}}$ and $\eta_{s}=\sqrt{\frac{1}{3}}\eta_{0}-\sqrt{\frac{2}{3}}\eta_{8}$.
The states of $\pi_{3}$, $\eta_{n}$ and $\eta_{s}$ are represented as $\pi_{3}=\frac{1}{\sqrt{2}}|u\bar{u}-d\bar{d}>$, $\eta_{n}=\frac{1}{\sqrt{2}}|u\bar{u}+d\bar{d}>$
and $\eta_{s}=|s\bar{s}>$ from the quark model, respectively.
The physical meson states $\pi^{0}$, $\eta$, $ \eta'$ can be transformed from the $\pi_{3}$, $\eta_{n}$ and $\eta_{s}$ by unitary matrix $U$ \cite{ Kroll2005PP}:
\begin{gather}
\label{Ubian}
\begin{pmatrix} \pi^{0}\\ \eta \\ \eta'\end{pmatrix}=U(\varepsilon_{1},\varepsilon_{2},\phi)\begin{pmatrix} \pi_{3}\\ \eta_{n}\\ \eta_{s} \end{pmatrix},
\end{gather}
where
\begin{gather}
U(\varepsilon_{1},\varepsilon_{2},\phi)=\begin{pmatrix} 1 & \varepsilon_{1}+\varepsilon_{2}cos\phi & -\varepsilon_{2}sin\phi \\
-\varepsilon_{2}-\varepsilon_{1}cos\phi & cos\phi & -sin\phi \\ -\varepsilon_{1}sin\phi & sin\phi & cos\phi \end{pmatrix},
\label{pm}
\end{gather}
where $\phi$ is the mixing angle \cite{AMLi2007}. $\varepsilon_{1}$, $\varepsilon_{2}$
$\propto \mathcal{O}(\lambda),\lambda \ll 1$ and the higher order terms are neglected.

From the quark model, the physical states $\eta$  and $\eta'$ are related to quark flavor basis $\eta_{n}$
and $\eta_{s}$. The matrix $U(\varepsilon_{1},\varepsilon_{2},\phi)$ reduces to the $U'(\phi)$ as $\varepsilon_{1},\varepsilon_{2}\rightarrow 0$ when one considers the isospin symmetry.
Hence, the physical state $\pi^{0}$ is equivalent to the $\pi_{3}$ state and
the formula of Eq.(\ref{Ubian})(\ref{pm}) translates into the $\eta-\eta'$ mixing in Eq.(\ref{1z}):
\begin{equation}
\left(\begin{matrix} \eta \\ \eta'
\end{matrix}
\right)=U'(\phi)\left(\begin{matrix} \eta_{n} \\ \eta_{s}
\end{matrix}
\right)
=
\left( \begin{array}{ccc}
\cos\phi & -sin\phi \\
\sin\phi  & cos\phi
\end{array} \right)
\left(\begin{matrix} \eta_{n} \\ \eta_{s}
\end{matrix}
\right).
\label{1z}
\end{equation}

The mechanism of isospin symmetry breaking depends on the electroweak interaction and the strong interaction by the $u-d$ quark mass difference in Standard Model.
For the isospin symmetry breaking, the effect of the electroweak interaction is tiny\cite{NG,MG,RF}.
The $u-d$ quark mass difference is responsible for $\pi^{0}-\eta-\eta'$ mixing
and make major contribution for the isospin symmetry breaking.
One can calculate the effects from isospin symmetry breaking by the chiral perturbative theory.
To the leading order of isospin symmetry breaking via $\pi^{0}-\eta-\eta'$ mixing,
the physical eigenstates of mesons
$\pi^{0}$, $\eta$ and $\eta'$ from Eq.(\ref{Ubian})(\ref{pm}) can be written as
\begin{eqnarray}
\begin{split}
|\pi^{0}\rangle&=|\pi_{3}\rangle+(\varepsilon_{1}+\varepsilon_{2}\cos\phi)|\eta_{n}\rangle-\varepsilon_{2}\sin\phi|\eta_{s}\rangle,\\
|\eta\rangle&=(-\varepsilon_{2}-\varepsilon_{1}\cos\phi)|\pi_{3}\rangle+\cos\phi|\eta_{n}\rangle-\sin\phi|\eta_{s}\rangle,\\
|\eta'\rangle&=-\varepsilon_{1}\sin\phi|\pi_{3}\rangle+\sin\phi|\eta_{n}\rangle+\cos\phi|\eta_{s}\rangle,
\end{split}
\label{zx}
\end{eqnarray}
where $\pi_{3}$  refer to the isospin $I=1$ component in the triplet. One can define $\varepsilon=\varepsilon_{2}+\varepsilon_{1}cos\phi$, $\varepsilon'=\varepsilon_{1}sin\phi$, and the values are $\varepsilon=0.017 \pm 0.002$ , $\varepsilon'=0.004 \pm 0.001$, $\phi=39.0^\circ$ \cite{Kroll2005PP}.

We need to deal with the hadronic matrix element to calculate the $CP$ violation.
Due to the high-energy of heavy meson, the factorization method is
widely used to estimate the hadronic matrix element
from the heavy quarks effective theory and the large momenta of quarks.
Because of the QCD correction, we can calculate the
decay amplitudes of the heavy $B$ meson since
soft gluons is not enough time to exchange between
the final states mesons with large momenta.
The contribution of hard gluons can be estimated
by perturbative theory.
In the framework of perturbative QCD,the transverse momenta of quark is considered to avoid the divergence.
One need pay close attention to three scales:
$m_{W}$ scale ($m_{W}$ refers to the W-boson mass),
the hard scale $t$, and factorization scale $1/b$
($b$ is the conjugate variable of the parton transverse momenta $k_{T}$).
The Wilson coefficients $C(t)$ is related to
contribution of short distance by the leading logarithm order.
One can resume the results from the order of $m_{W}$ down to hadronic scale $t$ through the renormalization group equation. The non-perturbative contribution can be absorbed into
hadronic wave function $\Phi(x)$ below factorization scale $k_{T}$.
The overlap of soft divergence and collinear divergence will result in
double logarithms $ln^{2}(Pb)$ ($P$ refers to the
light-cone component of meson momentum).
The resummation
of the double logarithms generates a Sudakov factor
$exp[-s(P,b)]$, which can suppress the long distance
contribution from the non-perturbative effects
in the large $b$ region, and vanishes
as $b>1/\Lambda_{QCD}$ fortunately.
We can calculate the remaining finite
subamplitude $H(x,t)$ by pertubative theory.
Hence, we present the $CP$ violation via $\pi^{0}-\eta-\eta'$ mixing
in the pertubative QCD scheme.

Since the wave function is independent on the decay mode, we use the model for the $B$ meson function\cite{AMLi2007,wangz2014}
\begin{eqnarray}
\phi_{B}(x,b)= N_{B}{x^2}(1-x)^2exp[-\frac {{M_{B}^2}x^2}{2{\omega_b}^2} - \frac 12 (\omega_b{b})^2],
\end{eqnarray}
where the normalization factor $N_{B}$ is dependent of the free parameter $\omega_b$.
$b$ is the conjugate variable of the parton transverse
momenta $k_{T}$. $M_{B}$ refers to the mass of the $B$ meson.
For the $B$ meson, we can obtain the value of $\omega_b = 0.50 \pm 0.05$
from the light cone sum rule \cite{hnl2002}.

The wave function of the final state pseudoscalar mesons $M=(\pi, K, \eta_{n}, \eta_{s})$ are the same in form and can be defined in Ref.\cite{VM2004,ZH2014,VM1990,YY2013}
\begin{eqnarray}
\Phi_{M_i}(P_{i},x_{i})\equiv\frac{1}{\sqrt{6}}\gamma_{5}[\not{P_{i}} \phi_{M_i}^{A}(x_{i})+ m_{0i}\phi_{M_i}^{P}(x_{i}) + \zeta(\not{n}\not{\upsilon}-1)m_{0i}\phi^{T}(x_{i}) ],
\end{eqnarray}
where $m_{0i}$ is the chiral mass of the meson $M_{i}$. $P_{i}$ and $x_{i}$ are the momentum and the fraction of the momentum of $M_{i}$. Depending on the assignment of the momentum fraction $x$, the parameter $\zeta$ should be chosen as +1 or -1.

In this paper, we will use those distribution amplitudes of the $\pi$ meson and $\phi_{\eta_{d\bar{d}}}^{A,P,T}$ \cite{Lu2006}:

\begin{eqnarray}
\begin{split}
 \phi_{\pi}^A(x) &=  \frac{3f_{\pi}}{\sqrt{6}} x(1-x)[ 1 +0.44C_2^{3/2}(2x-1)+0.25C_4^{3/2}(2x-1)],\\
 \phi_{\pi}^P(x) &=  \frac{f_{\pi}}{2\sqrt{6}}[1 +0.43C_2^{1/2}(2x-1)+ 0.09C_4^{1/2}(2x-1) ], \\
 \phi_{\pi}^T(x) &=  \frac{f_{\pi}}{2\sqrt{6}}(1-2x)[1+0.55(10x^2-10x+1) ] ,\\
 \phi_{\eta_{d\bar{d}}}^A(x) &= \frac{3}{{2\sqrt{2N_{c}}}}f_x x(1-x)[ 1 + a_{2}^{\eta_{d\bar{d}}}\frac{3}{2}(5(1-2x)^2-1)  \nonumber\\
  &+ a_{4}^{\eta_{d\bar{d}}}\frac{15}{8}(21(1-2x)^4-14(1-2x)^2-1)],\\
 \phi_{\eta_{d\bar{d}}}^P(x) &=  \frac{1}{2\sqrt{2N_c}} f_x [1 +\frac{1}{2}(30 \eta_3 - \frac{5}{2}\rho_{\eta_{d\bar{d}}}^{2})(3(1-2x)^2-1) \nonumber\\
 & +\frac{1}{8}(-3\eta_3 \omega_3 - \frac{27}{20}\rho_{\eta_{d\bar{d}}}^{2}  -\frac{81}{10}\rho_{\eta_{d\bar{d}}}^{2}a_{2}^{\eta_{d\bar{d}}})(35(1-2x)^4-30(1-2x)^2+3) ], \\
 \phi_{\eta_{d\bar{d}}}^T(x) &= \frac{3}{2\sqrt{2N_c}}f_x(1-2x) [\frac{1}{6}(5 \eta_3- \frac{1}{2}\eta_3 \omega_3- \frac{7}{20}\rho_{\eta_{d\bar{d}}}^{2}a_{2}^{\eta_{d\bar{d}}}- \frac{3}{5}\rho_{\eta_{d\bar{d}}}^{2}a_{2}^{\eta_{d\bar{d}}})(10x^2-10x+1) ] ,\\
\end{split}
\end{eqnarray}
where $\rho_{\eta_{d\bar{d}}}= {m_\pi}/{m_0^{\eta_{d\bar{d}}}}$. The distribution amplitudes $\phi_{\eta_{d\bar{d}}}^{A,P,T} $ represent the axial vector, pseudoscalar and tensor components of the wave function, respectively. In Ref.\cite{TF1998}, isospin symmetry is assumed for $f_x$ and $f_x=f_\pi$. The Gegenbauer polynomials are given as:
\begin{equation}
\begin{array}{ll}
 C^{1/2}_{2}(t)=\frac{1}{2}(3t^2-1) ,& C^{1/2}_{4}(t)=\frac{1}{8}(3-30t^2+35t^4)  \\
C^{3/2}_{1}(t)=3t,&C^{3/2}_{2}(t)=\frac{3}{2} (5t^2-1),
 \\
 C^{3/2}_{4}(t)=\frac{15}{8}(1-14t^2+21t^4).
\end{array}
\end{equation}
where $t=2x-1$.

The expressions of the distribution amplitudes of the $K$ meson and $\phi_{\eta_n}^{A,P,T}$ are given as \cite{X2008}:
\begin{eqnarray}
\begin{split}
 \phi_{K}^A(x) &= \frac{f_{K}}{{2\sqrt{2N_{c}}}} 6x(1-x)[ 1 + a_{1}^{K}C_1^{3/2}(t)+ + a_{2}^{K}C_2^{3/2}(t)+ + a_{4}^{K}C_4^{3/2}(t)],\\
 \phi_{K}^P(x) &=  \frac{f_{K}}{2\sqrt{2N_c}}[1 +(30 \eta_3 - \frac{5}{2}\rho_{K}^{2})C_2^{1/2}(t)-3(\eta_3 \omega_3 + \frac{9}{20}\rho_{K}^{2}(1 + 6 a_{2}^{K})) C_4^{1/2} ], \\
 \phi_{K}^T(x) &= -\frac{f_{K}}{2\sqrt{2N_c}}t [1 + 6(5 \eta_3- \frac{1}{2}\eta_3 \omega_3- \frac{7}{20}\rho_{K}^{2}- \frac{3}{5}\rho_{K}^{2}a_{2}^{K})(1+10x+10x^2) ] ,\\
 \phi_{\eta_n}^A(x) &= \frac{f_n}{{2\sqrt{2N_{c}}}} 6x(1-x)[ 1 + a_{1}^{\eta_n}C_1^{3/2}(2x-1)+ a_{2}^{\eta_n}C_2^{3/2}(2x-1)+ a_{4}^{\eta_n}C_4^{3/2}(2x-1)],\\
 \phi_{\eta_n}^P(x) &=  \frac{f_n}{2\sqrt{2N_c}}[1 +(30 \eta_3 - \frac{5}{2}\rho_{\eta_n}^{2})C_2^{1/2}(2x-1)-3(\eta_3 \omega_3 + \frac{9}{20}\rho_{\eta_n}^{2}(1 + 6 a_{2}^{\eta_n})) C_4^{1/2}(2x-1) ], \\
 \phi_{\eta_n}^T(x) &= \frac{f_n}{2\sqrt{2N_c}}(1-2x) [1 + 6(5 \eta_3- \frac{1}{2}\eta_3 \omega_3- \frac{7}{20}\rho_{\eta_n}^{2}- \frac{3}{5}\rho_{\eta_n}^{2}a_{2}^{\eta_n})(1+10x+10x^2) ] ,\\
 \end{split}
\end{eqnarray}
where $\rho_{K}=\frac{m_K}{m_{0K}}$ and $\rho_{\eta_n}=\frac{2 m_q}{m_{qq}}$ .The wave function of the $u\bar{u}$ is same as the wave function $d\bar{d}$. The Gegenbauer moments and the other parameters are given as \cite{X2008}:
 \begin{equation}
\begin{array}{ll}
a_{1}^{K}=0.06 ,& a_{2}^{K}=0.25,  \\
a_{4}^{K}=0,&a_{1}^{\eta_n}=0,
 \\
 a_{2}^{\eta_n, \eta_{d\bar{d}}}=0.44,& a_{4}^{\eta_n, \eta_{d\bar{d}}}=0.25  \\
 a_{2}^{\pi}=0.25 ,& a_{4}^{\pi}=-0.015,  \\
 \eta_3=0.015,& \omega_3=-3.0.
\end{array}
\end{equation}

The wave function of the meson is non-perturbative and process-independent.
We can describe the formation of hadrons from the positive and negative quarks by the wave function,
which provides the distribution of the momentum carried by the parton.
In the framework of PQCD, we calculate the QCD correction by introducing the transverse momentum.
Hence, the meson wave function should be dependent on transverse momentum.
The relevant results show that transverse momentum has large effects on the heavy meson function.
However, there is little effect on the wave function of light meson by transverse momentum.
At present, only the Light-cone QCD sum rule and Lattice QCD are credible and used to calculate non-perturbative contributions.
Generally, one obtains the wave function of optical pseudoscalar meson
by Light-cone QCD sum rule.Within the framework of PQCD, The branching ratios and CP violations have been calculated. The theoretical prediction is inconsistent with the experimental results for the most decay processes of $B/B_{s}\rightarrow M_{2}M_{3}$ \cite{LKM,X2008}. Hence, the wave function is credible \cite{ganggang2019}.

The relevant decay constants can be written as \cite{tpbm}:
\begin{eqnarray}
\begin{split}
\langle 0|\bar n\gamma^\mu\gamma_5 n|\eta_n(P)\rangle&= \frac{i}{\sqrt2}\,f_n\,P^\mu \;, \\
\langle 0|\bar s\gamma^\mu\gamma_5 s|\eta_s(P)\rangle &= i f_s\,P^\mu\;,
\end{split}
\label{deffq}
\end{eqnarray}
where $P$ refers to the momenta of $\eta_n$ or $\eta_s$. The relation between the decay constants  $f_n$ and $f_s$ can be found  in Ref.\cite{X2008}.

\subsection{\label{subsec:form}Calculation details}
In the framework of PQCD, we can calculate the $CP$ violation for the decay modes  ${B}\rightarrow \pi\pi$ and ${B}\rightarrow \pi K$ from the effects of isospin symmetry breaking mechanism via $\pi^{0}-\eta-\eta'$ mixing. The tree level amplitude $T$ and penguin level amplitude $P$
are obtained by perturbative theory. For the sake of simplification,
we take the decay process ${B}^{0}\rightarrow \pi^{0}\pi^{0}$  as example for showing the $\pi^{0}-\eta-\eta'$ mixing mechanism.

The decay amplitudes $\mathcal{A}$ for ${B}^{0}\rightarrow \pi^{0}\pi^{0}$
via $\pi^{0}-\eta-\eta'$ mixing can be written as :
\begin{eqnarray}
\mathcal{A}&=\langle\pi^{0}\pi^{0}|{\mathcal{H}}_{eff}|B^{0}
\rangle={\mathcal{A}}_1 + {\mathcal{A}}_2 + {\mathcal{A}}_3
\label{A}
\end{eqnarray}
where
\begin{eqnarray}
\begin{split}
{\mathcal{A}}_1&= \langle\pi_3\pi^{0}|{\mathcal{H}}_{eff}|B^{0}\rangle \\
& =\langle\pi_3\pi_3|{\mathcal{H}}_{eff}|B^0 \rangle+(\varepsilon_1+\varepsilon_2\cos\phi)\langle\pi_3\eta_{n}|{\mathcal{H}}_{eff}|B^0\rangle-\varepsilon_2\sin\phi
\langle\pi_3\eta_{s}|{\mathcal{H}}_{eff}|B^0\rangle,
\end{split}
\label{ma1}
\end{eqnarray}
\begin{eqnarray}
\begin{split}
{\mathcal{A}}_2& =(\varepsilon_1+\varepsilon_2\cos\phi)\langle\pi^{0}\eta_n|{\mathcal{H}}_{eff}|B^{0}\rangle  \\
&  =(\varepsilon_1+\varepsilon_2\cos\phi)\langle\pi_3\eta_n|{\mathcal{H}}_{eff}|B^{0}\rangle+\mathcal{O} (\varepsilon),
\end{split}
\label{ma2}
\end{eqnarray}
and
\begin{eqnarray}
\begin{split}
{\mathcal{A}}_3& =-\varepsilon_{2}\sin\phi\langle\pi^{0}\eta_{s}|{\mathcal{H}}_{eff}|B^0\rangle  \\
&
= -\varepsilon_{2}\sin\phi\langle\pi_3\eta_{s}|{\mathcal{H}}_{eff}|B^{0}\rangle+\mathcal{O} (\varepsilon),
\end{split}
\label{ma3}
\end{eqnarray}
and we have ignored the higher order term of $\varepsilon$ and $\mathcal{O} (\varepsilon)=\mathcal{O} (\varepsilon_1)+\mathcal{O} (\varepsilon_2)$.

We obtained the amplitude forms required in Eq.(\ref{ma1})(\ref{ma2})(\ref{ma3}) by calculation and they can be written as
\begin{eqnarray}
\langle\pi_3\pi_3|{\mathcal{H}}_{eff}|B^{0}\rangle&=& -\frac{1}{\varepsilon_1}\Big\{ F_{e\pi}\left[\xi_u a_2- \xi_t(2a_3+a_4-2a_5-\frac{1}{2}a_7+\frac{1}{2}a_9-\frac{1}{2}a_{10}) f_\eta^d  -\xi_t(a_3-a_5+\frac{1}{2}a_7-\frac{1}{2}a_9) f_\eta^s\right]\nonumber\\
  &&
  + F_{e\pi}^{P2}  \xi_t(a_6-\frac{1}{2}a_8) f_\eta^d\Big\},
\end{eqnarray}

\begin{eqnarray}
\langle\pi_3\eta_{n}|{\mathcal{H}}_{eff}|B^{0}\rangle&=&\frac{1}{\sqrt{2}} \Big\{F_{e\eta} \left[\xi_u a_2- \xi_t(-a_4-\frac{3}{2}a_7+\frac{3}{2}a_9+\frac{1}{2}a_{10})\right] f_{\pi}+(M_{a\eta}^{P1}+M_{a\pi}^{P1})(C_5-\frac{1}{2}C_7)\nonumber\\
  &&
  + M_{e\pi}^{P2}\xi_t \left[(2C_{6}+\frac{1}{2}C_8)\right]+  (M_{a\pi}+M_{e\eta}+M_{a\eta})\left[\xi_u C_2- \xi_t(-C_3+\frac{1}{2}C_9+\frac{3}{2}C_{10})\right]\nonumber\\
  &&
 -\frac{3}{2}(M_{a\pi}^{P2}+M_{e\eta}^{P2}+M_{a\eta}^{P2})\xi_t C_8 + F_{e\pi}^{P2}\xi_t(a_6-\frac{1}{2}a_8)\Big\}\nonumber\\
  &&
 +\frac{\varepsilon_1}{\varepsilon_2}\Big\{ F_{e\pi}\left[\xi_u a_2- \xi_t(2a_3+a_4-2a_5-\frac{1}{2}a_7+\frac{1}{2}a_9-\frac{1}{2}a_{10}) f_\eta^d  -\xi_t(a_3-a_5+\frac{1}{2}a_7-\frac{1}{2}a_9) f_\eta^s\right]\nonumber\\
  &&
  + F_{e\pi}^{P2}  \xi_t(a_6-\frac{1}{2}a_8) f_\eta^d\Big\}+\frac{\varepsilon_1}{2\varepsilon_2}M_{e\pi} \left[\xi_u C_2- \xi_t(C_3+C_4-\frac{1}{2}C_9+\frac{1}{2}C_{10})\right],
\end{eqnarray}
and
\begin{eqnarray}
\langle\pi_3\eta_{s}|{\mathcal{H}}_{eff}|B^{0}\rangle&=&M_{e\pi}^{P2}\xi_t (2C_{6}+\frac{1}{2}C_8)+\frac{\varepsilon_1}{2\sqrt{2}\varepsilon_2}M_{e\pi} \left[-\xi_t(C_4-\frac{1}{2}C_{10})\right].
\end{eqnarray}

Depending on the CKM matrix elements, we can express the decay amplitudes $\mathcal{A}$ for ${B}^{0}\rightarrow \pi^{0}\pi^{0}$ as following :
\begin{eqnarray}
\begin{split}
 -\sqrt{2}\mathcal{A}({B}^{0}\rightarrow \pi^{0}\pi^{0})= \xi_u T-\xi_t P
\end{split}
\end{eqnarray}
where $\xi_u={V_{ub}^*}{V_{ud}}$, $\xi_t={V_{tb}^*}{V_{td}}$. $T$ and $P$ refer to the tree and penguin contributions from $\mathcal{A}$ in Eq.(\ref{A}), respectively.
The amplitudes T and P from the decay process of ${B}^{0}\rightarrow \pi^{0}\pi^{0}$ with $\pi^{0}-\eta-\eta'$ mixing
can be written as:
\begin{eqnarray}
\begin{split}
 T &= T_1 + (\varepsilon_{1}+\varepsilon_{2}\cos\phi)T_n-\varepsilon_{2}\sin\phi T_s,\\
 P &= P_1 + (\varepsilon_{1}+\varepsilon_{2}\cos\phi)P_n-\varepsilon_{2}\sin\phi P_s.
 \label{tp}
\end{split}
\end{eqnarray}
where $T_1$, $P_1$ are from the decay process of ${B}^{0}\rightarrow \pi^{0}\pi^{0}$ without $\pi^{0}-\eta-\eta'$ mixing. $T_n$, $P_n$, $T_s$ and $P_s$ come from the decay amplitudes ${B}^{0}\rightarrow \pi^{0}\eta_n$ and ${B}^{0}\rightarrow \pi^{0}\eta_s$, respectively. The tree level amplitude and penguin level amplitude can be given as
\begin{eqnarray}
T_1=f_{\pi} F_e \left[c_{1}+\frac{1}{3}c_{1}\right]+ M_e [C_{2}]- M_a [C_{2}],
\label{t1}
\end{eqnarray}

\begin{eqnarray}
P_1&=&f_{\pi} F_e \left[
 \frac{1}{3}C_{3}+C_4+\frac{3}{2}C_7+\frac{1}{2}C_8-\frac{5}{3}C_9-C_{10}\right]+f_{\pi} F_e^{P} \left[
 C_6+\frac{1}{3}C_5-\frac{1}{6}C_7-\frac{1}{2}C_8\right] \nonumber\\
  &&
 -  M_e \left[-C_{3}+\frac{3}{2}C_8+\frac{1}{2}C_{9}+\frac{3}{2}C_{10}\right]
  +  M_e\left[C_{3}+2C_4++2C_6+\frac{1}{2}C_{8}-\frac{1}{2}C_{9}+\frac{1}{2}C_{10}\right]\nonumber\\
  &&
  +  f_{B} F_a \left[
 \frac{1}{3}C_{5}+C_6-\frac{1}{6}C_7-\frac{1}{2}C_8\right],
\label{p1}
\end{eqnarray}

\begin{eqnarray}
T_n&=&\frac{1}{\sqrt{2}} [F_{e\eta} f_{\pi}a_2 + (M_{a\pi}+M_{e\eta}+M_{a\eta}) C_2]
 +\frac{\varepsilon_1}{\varepsilon_2} F_{e\pi} a_2+\frac{\varepsilon_1}{2\varepsilon_2}M_{e\pi}  C_2,
\end{eqnarray}

\begin{eqnarray}
 P_n&=&\frac{1}{\sqrt{2}} \Big\{F_{e\eta} \left[- (-a_4-\frac{3}{2}a_7+\frac{3}{2}a_9+\frac{1}{2}a_{10})\right] f_{\pi}
  + M_{e\pi}^{P2} \left[(2C_{6}+\frac{1}{2}C_8)\right]\nonumber\\
  &&+  (M_{a\pi}+M_{e\eta}+M_{a\eta})\left[- (-C_3+\frac{1}{2}C_9+\frac{3}{2}C_{10})\right]
 -\frac{3}{2}(M_{a\pi}^{P2}+M_{e\eta}^{P2}+M_{a\eta}^{P2}) C_8 + F_{e\pi}^{P2}\xi_t(a_6-\frac{1}{2}a_8)\Big\}\nonumber\\
  &&
 +\frac{\varepsilon_1}{\varepsilon_2}\Big\{ F_{e\pi}\left[-(2a_3+a_4-2a_5-\frac{1}{2}a_7+\frac{1}{2}a_9-\frac{1}{2}a_{10}) f_\eta^d  -(a_3-a_5+\frac{1}{2}a_7-\frac{1}{2}a_9) f_\eta^s\right]\nonumber\\
  &&
  + F_{e\pi}^{P2}  (a_6-\frac{1}{2}a_8) f_\eta^d\Big\}+\frac{\varepsilon_1}{2\varepsilon_2}M_{e\pi} \left[- (C_3+C_4-\frac{1}{2}C_9+\frac{1}{2}C_{10})\right],
\end{eqnarray}

\begin{eqnarray}
\begin{split}
 T_s=0,
\end{split}
\end{eqnarray}
and
\begin{eqnarray}
\begin{split}
 P_s&=M_{e\pi}^{P2} (2C_{6}+\frac{1}{2}C_8)+\frac{\varepsilon_1}{2\sqrt{2}\varepsilon_2}M_{e\pi} \left[-(C_4-\frac{1}{2}C_{10})\right],
\end{split}
\end{eqnarray}
 with the $f_{\pi}$ and $f_{B}$ refer to the decay constant. The individual decay amplitudes in the above equations, such as $F_e$, $F_e^{P}$,  $F_a$,  $M_a$ and $M_e$ arise from the $(V-A)(V-A)$, $(V-A)(V+A)$ and $(S-P)(S+P)$ operators, respectively, and will be given in Appendix.

One can see that the Eq.(\ref{tp}) without $\pi^{0}-\eta-\eta'$ mixing is reduced to
\begin{eqnarray}
 T = T_1,\;\;\;\;\;\;\; P = P_1,
 \end{eqnarray}
which are expressed in Eq.(\ref{t1}) and Eq.(\ref{p1}).

The relevant weak phase $\theta$ and strong phase $\delta$ are obtained as following
\begin{eqnarray}
r e^{i\delta} e^{i\theta}&=&\frac{P}{T}
\times\frac{V_{td}V_{tb}^*}{V_{ud}V_{ub}^*},  \label{eq:delform}
\end{eqnarray}
where the parameter $r$ represents the
absolute value of the ratio of penguin and tree amplitudes:
\begin{eqnarray}
r\equiv\Bigg|\frac{\langle \pi^{0}\pi^{0}|H^P|{B}^{0}\rangle}{\langle \pi^{0}\pi^{0}|H^T|\bar{B}_{s}^{0}\rangle}\Bigg|
\label{r}.
\end{eqnarray}
The strong phase associated with $r$ can be given
\begin{eqnarray}
r e^{i\delta} &=&\frac{P}{T}
\times\bigg|\frac{V_{td}V_{tb}^*}{V_{ud}V_{ub}^*}\bigg| =r\cos\delta + \mathrm{i}r\sin\delta,  \label{eq:delform}
\end{eqnarray}
where
\begin{equation}
\left|\frac{V_{td}V^{*}_{tb}}{V_{ud}V^{*}_{ub}}\right|=\frac{\sqrt{[\rho(1-\rho)-\eta^2]^2+\eta^2}}{(1-\lambda^2/2)(\rho^2+\eta^2)}.
\label{3p}
\vspace{2mm}
\end{equation}
where $\rho$, $\eta$, $\lambda$ is the Wolfenstein parameters.

The $CP$ violation, $A_{CP}$, can be written as
\begin{eqnarray}
A_{CP}\equiv\frac{|A|^2-|\bar{A}|^2}{|A|^2+|\bar{A}|^2}=\frac{-2r
{\rm{sin}}\delta {\rm{sin}}\theta}{1+2r {\rm{cos}}\delta
{\rm{cos}}\theta+r^2}. \label{asy}
\end{eqnarray}

\section{\label{input}INPUT PARAMETERS}
The CKM matrix, which elements are determined from experiments, can be expressed in terms of the Wolfenstein parameters $A$, $\rho$, $\lambda$ and $\eta$ \cite{wol}:
\begin{equation}
\left(
\begin{array}{ccc}
  1-\tfrac{1}{2}\lambda^2   & \lambda                  &A\lambda^3(\rho-\mathrm{i}\eta) \\
  -\lambda                 & 1-\tfrac{1}{2}\lambda^2   &A\lambda^2 \\
  A\lambda^3(1-\rho-\mathrm{i}\eta) & -A\lambda^2              &1\\
\end{array}
\right),\label{ckm}
\end{equation}
where $\mathcal{O} (\lambda^{4})$ corrections are neglected. The latest values for the parameters in the CKM matrix are \cite{PDG2018}:
\begin{eqnarray}
&& \lambda=0.22506\pm0.00050,\quad A=0.811\pm0.026,\nonumber \\
&& \bar{\rho}=0.124_{-0.018}^{+0.019},\quad
\bar{\eta}=0.356\pm{0.011}.\label{eq: rhobarvalue}
\end{eqnarray}
where
\begin{eqnarray}
 \bar{\rho}=\rho(1-\frac{\lambda^2}{2}),\quad
\bar{\eta}=\eta(1-\frac{\lambda^2}{2}).\label{eq: rho rhobar
relation}
\end{eqnarray}
From Eqs. (\ref{eq: rhobarvalue}) ( \ref{eq: rho rhobar relation})
we have
\begin{eqnarray}
0.109<\rho<0.147,\quad  0.354<\eta<0.377.\label{eq: rho value}
\end{eqnarray}
The other parameters are given as following \cite{Lu2006,xiao2007,PDG2018}:
\begin{eqnarray}
f_{\pi}&=&0.13\text{GeV}, \hspace{2.95cm}  f_{B}=0.19\text{GeV},        \nonumber \\
m_{B}&=&5.2792\text{GeV},\hspace{2.65cm} f_{K}=0.16\text{GeV},   \nonumber \\
f_s&=&0.17\text{GeV},\hspace{2.95cm} f_n=0.14\text{GeV},  \nonumber \\
m_\pi&=&0.14\text{GeV},\hspace{2.95cm} m_W=80.41\text{GeV},  \nonumber \\
m_0^{\eta_{s\bar{s}}}&=&2.4\text{GeV},\hspace{2.95cm} m_0^{\eta_{d\bar{d}}}=1.4\text{MeV}.
\end{eqnarray}

\section{\label{num}Numerical results}

 The $CP$ violation depends on the weak phase differences from the CKM matrix elements and the strong phase differences which comes from the dynamics of QCD and the other mechanism. The CKM matrix elements are related to $A$, $\rho$, $\eta$ and $\lambda$, but the results for the $CP$ violation are less reliant on $A$ and $\lambda$ in the course of calculations. Hence, we present the $CP$ violation from the weak phases associated with the $\rho$ and $\eta$ in the CKM matrix while the $A$ and $\lambda$ are assigned for the central values.  The major uncertainties for the $CP$ violations is from the parameters, which arises from the uncertainties of parameter $\omega_{b} =0.50\pm0.05Gev$ the variations of $m_0^K=1.6\pm0.1Gev$, $m_0^\pi=1.4\pm0.1Gev$ and the Gegenbauer Coefficients $a_2^{K}=0.25\pm0.15Gev$, $a_2^\eta=0.44\pm0.22Gev$. From Table.I, we can find that the isospin breaking via $\pi^{0}-\eta-\eta'$ mixing changes the sign of the $CP$ violation for the decay channel of ${B}^{0}\rightarrow \pi^{0}\pi^{0}$ and $B^0\rightarrow K^0\pi^{0}$, which compared data from three other methods for the central value. Compared with the PQCD, one can find that the central value of the $CP$ violation via $\pi^{0}-\eta-\eta'$ mixing has changed $0.25$ for the decay channel of ${B}^{0}\rightarrow \pi^{0}\pi^{0}$. This is due to the breaking of isospin symmetry, the interference between the $\pi^0$ and $\eta(')$ mesons is stronger than other decay modes. And there is twice contribution from $\pi^{0}-\eta-\eta'$ mixing.
We can find the $CP$ violation of the decay mode ${B}^{+}\rightarrow \pi^{+}\pi^{0}$ agree well with the QCDF and PQCD predictions in Table.I. For the decay mode ${B}^{+}\rightarrow K^{+}\pi^{0}$,
our result presents the large region for the $CP$ violation,
and include the experiment data. In comparison with the results of PQCD without
$\pi^{0}-\eta-\eta'$ mixing, the isospin symmetry breaking generates large effect.

\tabcolsep 0.29in
\begin{table}
\centering
\caption{The CP violation of $B$ decay mode via isospin symmetry breaking via $\pi^{0}-\eta-\eta^{'}$ mixing in PQCD.}
\begin{tabular}{ccccc}
\hline\hline \\
\multicolumn{1}{c}{decay mode} & QCDF\cite{MB1} &PQCD\cite{lhn2005} & $\pi^{0}-\eta-\eta^{'}$ mixing& Data\cite{HFAG,MP2014}\\ \\
\hline
$B^0\rightarrow\pi^0\pi^0$& $0.451^{+0.184}_{-0.028}$& $0.07\pm 0.03$&$-0.18^{+0.19}_{-0.01}$& $0.03\pm 0.17$  \\
$B^0\rightarrow\pi^+\pi^0$&$0.000^{+0.000}_{-0.000}$& $0.00\pm 0.00$& $0.00^{+0.00}_{-0.00}$& $0.026\pm 0.039$ \\
$B^0\rightarrow K^0\pi^0$& $-0.033^{+0.010}_{-0.008}$& $-0.09^{+0.06}_{-0.08}$& $0.03^{+0.00}_{-0.04}$& $-0.01\pm 0.10$  \\
$B^0\rightarrow K^+\pi^0$& $0.071^{+0.017}_{-0.018}$& $-0.01^{+0.03}_{-0.05}$& $-0.45^{+0.63}_{-0.03}$& $0.040\pm 0.021$  \\
\hline\hline
\end{tabular}
\end{table}

\section{\label{sum}SUMMARY AND DISCUSSION}
In this paper, we study the $CP$ violation for the ${B}^{0}\rightarrow \pi^{0}\pi^{0}$, ${B}^{+}\rightarrow \pi^{+}\pi^{0}$, ${B}^{0}\rightarrow K^{0}\pi^{0}$ and ${B}^{+}\rightarrow K^{+}\pi^{0}$ decays from isospin symmetry breaking effects via $\pi^{0}-\eta-\eta'$
in perturbative QCD. It is found that the $CP$ violation can be changed  via $\pi^{0}-\eta-\eta'$ mixing from the isospin symmetry breaking from the strong phase difference.

The traditional factorization method is the leading order of QCD factorization scheme with regard to the decay processes of bottom mesons and $1/m_b$ corrections is ignored.
The QCD factorization scheme is affected by the singularities of endpoint, which are not well controlled.
Recently, next-to-leading order corrections for the decay of
$B\rightarrow\pi\rho(\omega)$ have  been implemented
in the PQCD approach \cite{lu-next}. Some questions of the NLO accurate calculation in
PQCD approach are discussed by the author in article \cite{beneke-proc}. The complete NLO calculation in the PQCD method
needs to compute all one-loop spectator-scattering diagrams.
Nevertheless, the paper \cite{lu-next} only takes into account  the 1-loop BBNS kernel
(similar to the kernel in QCD factorization) for the decay of
$B\rightarrow\pi\rho(\omega)$. The vertex diagram can be far away from the shell as a subdiagram of a large diagram with hard collinear commutation. Besides, the Wilson coefficients are evaluated at a very low scale where the perturbation theory is destroyed.
The non-physical strengthen of the Wilson coefficients at small scales is also the source of the
large penguin and annihilation.
Based on above discussion, we
investigate the $CP$ violation of $B \rightarrow \pi\pi(K)$ via $\pi^{0}-\eta-\eta'$  mixing
from the leading order contribution in PQCD.

Theoretical errors lead to the uncertainty of results. In general,the power corrections in the heavy quark limit will give the main theoretical uncertainty.
It means that introduce $1/m_b$ power corrections is essential.
Unluckily, there are many possibilities of $1/m_b$ power inhibiting effects and they are nonperturbative essentially, so they unable to calculate by the perturbative method. This scheme has more uncertainty. The first error relates to the change in the CKM parameters. The second error comes from hadron parameters, that is, the shape parameter, shape factor, decay constant, and the wave function of $B$ meson. The third error corresponds to the selection of the hard scales, which ranges from 0.75t to 1.25t, indicating the size of next-to-leading order in QCD contributions.

\section{Acknowledgments}
This work was supported by National Natural Science
Foundation of China (Project Numbers 11605041), Excellent Thesis Cultivation Project of Henan University of Technology
and Science and Education Integration Project of Henan University of Technology.

\section{APPENDIX: Related functions defined in the text}
The functions related with the tree and penguin contributions are presented  with PQCD approach \cite{LKM,AMLi2007,YK2001}.

The hard scales $t$ are chosen as \begin{eqnarray}
t_a&=&\mbox{max}\{{\sqrt{x_2}m_{B},1/b_2,1/b_3}\},\\
t_e^1&=&\mbox{max}\{{\sqrt{x_3}m_{B},1/b_1,1/b_3}\},\\
t_e^2&=&\mbox{max}\{{\sqrt{x_1}m_{B},1/b_1,1/b_3}\},\\
t_e^3&=&\mbox{max}\{\sqrt{x_3}m_{B},1/b_2,1/b_3\},\\
t_e^4&=&\mbox{max}\{\sqrt{x_2}m_{B},1/b_2,1/b_3\},\\
t_f&=&\mbox{max}\{\sqrt{x_1x_3}m_{B},\sqrt{(x_1-x_2)x_3}m_{B},1/b_1,1/b_2\},\\
t_f^1&=&\mbox{max}\{{\sqrt{x_2x_3}m_{B},1/b_1,1/b_2}\},\\
t_f^2&=&\mbox{max}\{{\sqrt{x_2x_3}m_{B},\sqrt{x_2+x_3-x_2x_3}m_{B},1/b_1,1/b_2}\},\\
t_f^3&=&\mbox{max}\{\sqrt{x_1+x_2+x_3-x_1x_3-x_2x_3}m_{B},\sqrt{x_2x_3}m_{B},1/b_1,1/b_2\},\\
t_f^4&=&\mbox{max}\{\sqrt{x_2x_3}m_{B},\sqrt{(x_1-x_2)x_3}m_{B},1/b_1,1/b_2\}.
\end{eqnarray}

The function $h$ coming from the Fourier transformations of the function $H^{(0)}$ \cite{xiao2007}.
They are defined by
\begin{eqnarray}
h_e(x_1,x_3,b_1,b_3)&=&K_0(\sqrt {x_1x_3}m_{B}b_1밀\left[\theta(b_1-b_3)K_0(\sqrt x_3 m_{B}b_1)I_0(\sqrt x_3m_{B}b_3)\right.\\
&& \left.+\theta(b_3-b_1)K_0(\sqrt x_3m_{B}b_3)I_0(\sqrt x_3m_{B}b_1)\right]S_t(x_3),\nonumber
\\
h_e^1(x_1,x_2,b_1,b_2)&=&K_0(\sqrt {x_1x_2}m_{B}b_1밀\left[\theta(b_1-b_2)K_0(\sqrt x_2 m_{B}b_1)I_0(\sqrt x_2m_{B}b_2)\right.\\
&& \left.+\theta(b_2-b_1)K_0(\sqrt x_2m_{B}b_2)I_0(\sqrt x_2m_{B}b_1)\right],\nonumber
\\
h_a(x_2,x_3,b_2,b_3)&=&K_0(i\sqrt {x_2x_3}m_{B}b_2밀\left[\theta(b_3-b_2)K_0(i\sqrt x_3 m_{B}b_3)I_0(i\sqrt x_3m_{B}b_2)\right.\\
&& \left.+\theta(b_2-b_3)K_0(i\sqrt x_3m_{B}b_2)I_0(i\sqrt x_3m_{B}b_3)\right]S_t(x_3),\nonumber
\\
h_d(x_1,x_2,x_3,b_1,b_2)&=&K_0(-i\sqrt {x_2x_3}m_{B}b_2밀\left[\theta(b_1-b_2)K_0(\sqrt x_1x_2 m_{B}b_1)I_0(\sqrt x_1x_2m_{B}b_2)\right.\\
&& \left.+\theta(b_2-b_1)K_0(\sqrt x_1x_2m_{B}b_2)I_0(\sqrt x_1x_2m_{B}b_1)\right],\nonumber
\\
h_f(x_1,x_2,x_3,b_1,b_2)&=&[\theta(b_2-b_1)I_0(M_{B}\sqrt{x_1x_3}b_1)K_0(M_{B}\sqrt{x_1x_3}b_2) \\
&&+(b_1\longleftrightarrow b_2)]\cdot
\left\{\begin{array}{ll}K_0(M_{B}F_{(1)}b_2),& for \;\;F_{(1)}^2>0\\
\frac{i\pi}{2} H_0^{(1)} (M_{B}\sqrt{|F_{(1)^2}|}b_2),& for \;\;F_{(1)}^2<0
\end{array}
\right. ,  \nonumber
\\
h_f^1(x_1,x_2,x_3,b_1,b_2)&=&K_0(-i\sqrt {x_2x_3}m_{B}b_1밀\left[\theta(b_1-b_2)K_0(-i\sqrt x_2x_3 m_{B}b_1)J_0(\sqrt x_2x_3m_{B}b_2)\right.\\
&& \left.+\theta(b_2-b_1)K_0(-i\sqrt x_2x_3m_{B}b_2)J_0(\sqrt x_2x_3m_{B}b_3)\right],\nonumber
\\
h_f^2(x_1,x_2,x_3,b_1,b_2)&=&K_0(i\sqrt {x_2+x_3-x_2x_3}m_{B}b_1밀\left[\theta(b_1-b_2)K_0(-i\sqrt x_2x_3 m_{B}b_1)J_0(\sqrt x_2x_3m_{B}b_2)\right.\\
&& \left.+\theta(b_2-b_1)K_0(-i\sqrt x_2x_3m_{B}b_2)J_0(\sqrt x_2x_3m_{B}b_1)\right],\nonumber
\\
h_f^3(x_1,x_2,x_3,b_1,b_2)&=&[\theta(b_1-b_2)K_0(i\sqrt{x_2x_3}b_1M_{B})I_0(i\sqrt{x_2x_3}b_2M_{B}) \\
&&+(b_1\longleftrightarrow b_2)]\cdot K_0(\sqrt{x_1+ x_2+ x_3- x_1x_3- x_2x_3}b_1M_{B})\nonumber
\\
h_f^4(x_1,x_2,x_3,b_1,b_2)&=&[\theta(b_1-b_2)K_0(i\sqrt{x_2x_3}b_1M_{B})I_0(i\sqrt{x_2x_3}b_2M_{B}) \\
&&+(b_1\longleftrightarrow b_2)]\cdot
\left\{\begin{array}{ll}K_0(M_{B}F_{(2)}b_2),& for \;\;F_{(2)}^2>0\\
\frac{i\pi}{2} H_0^{(1)} (M_{B}\sqrt{|F_{(2)^2}|}b_2),& for \;\;F_{(2)}^2<0
\end{array}
\right. ,
\end{eqnarray}
where $J_0$ is the Bessel function and $K_0$, $I_0$ are the modified Bessel functions $K_0(-ix) = -\frac{\pi}{2}\mathrm{y}_0(x) + i\,\frac{\pi}{2} \mathrm{J}_0(x)$, and $F_{(j)}$'s are defined by
\begin{eqnarray}
F_{(1)}^2=(x_1-x_2)x_3, \;\;\;\;  F_{(2)}^2=(x_1-x_2)x_3.
\end{eqnarray}

The $S_t$ re-sums the threshold logarithms $\ln^2x$ appearing in the
hard kernels to all orders and it has been parameterized as
\begin{eqnarray}
S_t(x)=\frac{2^{1+2c}\Gamma(3/2+c)}{\sqrt \pi
\Gamma(1+c)}[x(1-x)]^c,
\end{eqnarray}
with $c=0.3$. In the nonfactorizable contributions, $S_t(x)$ gives
a very small numerical effect on the amplitude~\cite{L4}.

The Sudakov exponents are defined as
\begin{eqnarray}
S_{B}(t)&=&s\left(x_1\frac{m_{B}}{\sqrt2},b_1\right)-\frac{1}{\beta_1}\ln\frac{\ln(t/\Lambda)}{-\ln(b_1\Lambda)},\\
S_{\pi}^1(t)&=&s\left(x_2\frac{m_{B}}{\sqrt2},b_2\right)+s\left((1-x_2)\frac{m_{B}}{\sqrt2},b_2\right)
    -\frac{1}{\beta_1}\ln\frac{\ln(t/\Lambda)}{-\ln(b_2\Lambda)},\\
S_{\pi}^2(t)&=&s\left(x_3\frac{m_{B}}{\sqrt2},b_3\right)+s\left((1-x_3)\frac{m_{B}}{\sqrt2},b_3\right)
      -\frac{1}{\beta_1}\ln\frac{\ln(t/\Lambda)}{-\ln(b_3\Lambda)},\\
S_{ab}(t)&=&s\left(x_1\frac{m_{B}}{\sqrt2},b_1\right)+s\left(x_3\frac{m_{B}}{\sqrt2},b_3\right)+s\left((1-x_3)\frac{m_{B}}{\sqrt2},b_3\right)
   -\frac{1}{\beta_1}\left[\ln\frac{\ln(t/\Lambda)}{-\ln(b_1\Lambda)}+\ln\frac{\ln(t/\Lambda)}{-\ln(b_3\Lambda)}\right],\\
S_{cd}(t)&=&s\left(x_1\frac{m_{B}}{\sqrt2},b_1\right)+s\left(x_2\frac{m_{B}}{\sqrt2},b_2\right)+s\left((1-x_2)\frac{m_{B}}{\sqrt2},b_2\right)
     +s\left(x_3\frac{m_{B}}{\sqrt2},b_1\right)+s\left((1-x_3)\frac{m_{B}}{\sqrt2},b_1\right)\\
&&
     -\frac{1}{\beta_1}\left[\ln\frac{\ln(t/\Lambda)}{-\ln(b_1\Lambda)}+\ln\frac{\ln(t/\Lambda)}{-\ln(b_2\Lambda)}\right],\\
S_{ef}(t)&=&s\left(x_1\frac{m_{B}}{\sqrt2},b_1\right)+s\left(x_2\frac{m_{B}}{\sqrt2},b_2\right)+s\left((1-x_2)\frac{m_{B}}{\sqrt2},b_2\right)
     +s\left(x_3\frac{m_{B}}{\sqrt2},b_2\right)+s\left((1-x_3)\frac{m_{B}}{\sqrt2},b_2\right)\\
&&   -\frac{1}{\beta_1}\left[\ln\frac{\ln(t/\Lambda)}{-\ln(b_1\Lambda)}+2\ln\frac{\ln(t/\Lambda)}{-\ln(b_2\Lambda)}\right],\\
S_{gh}(t)&=&s\left(x_2\frac{m_{B}}{\sqrt2},b_2\right)+s\left(x_3\frac{m_{B}}{\sqrt2},b_3\right)+s\left((1-x_2)\frac{m_{B}}{\sqrt2},b_2\right)
       +s\left((1-x_3)\frac{m_{B}}{\sqrt2},b_3\right)\\
&&    -\frac{1}{\beta_1}\left[\ln\frac{\ln(t/\Lambda)}{-\ln(b_2\Lambda)}+\ln\frac{\ln(t/\Lambda)}{-\ln(b_3\Lambda)}\right].\\
\end{eqnarray}
The explicit form for the  function
$s(k,b)$ is \cite{LKM}:
\begin{eqnarray}
s(k,b)&=&\frac{2}{3\beta_1}\left[\hat{q}\ln\left(\frac{\hat{q}}
{\hat{b}}-\hat{q}+\hat{b}\right)\right]+\frac{A^{(2)}}{4\beta_{1}^{2}}\left(\frac{\hat{q}}{\hat{b}}-1\right)\\
&&-\left[\frac{A^{(2)}}{4\beta_{1}^{2}}-\frac{1}{3\beta_{1}}(2\gamma_E-1-\ln2)\right]\ln\left(\frac{\hat{q}}{\hat{b}}\right),
\end{eqnarray}

where the variables are defined by
\begin{eqnarray}
\hat q\equiv \mbox{ln}[k/(\sqrt \Lambda)],~~~ \hat b\equiv
\mbox{ln}[1/(b\Lambda)], \end{eqnarray} and the coefficients
$A^{(i)}$ and $\beta_i$ are
\begin{eqnarray}
\beta_1&=&\frac{33-2n_f}{12},\\
A^{(2)}&=&\frac{67}{9}
-\frac{\pi^2}{3}-\frac{10}{27}n_f+\frac{8}{3}\beta_1\mbox{ln}(\frac{1}{2}e^{\gamma_E}),
\end{eqnarray}
$n_f$ is the number of the quark flavors and $\gamma_E$ is the
Euler constant.

The individual decay amplitudes  $F_e$, $F_e^{P}$,  $F_a$,  $M_a$ and $M_e$ arise from the $(V-A)(V-A)$ and $(V-A)(V+A)$ operators, respectively. The sum of their amplitudes are
\begin{itemize}
\item $(V-A)(V-A)$ operators:
 \begin{eqnarray}
  F_{e}&=&-16\pi C_Fm_{B}^4\int^1_0dx_1dx_2\int^\infty_0b_1db_1b_2db_2
\phi_{B}(x_1,b_1)\alpha_s(t_e^1)h_e^1(x_1,x_2,b_1,b_2)\nonumber\\
  &&\times\Big\{
   \Big[(1+x_2)\phi_\pi(x_2,b_2)+(1-2x_2)r_\pi\phi'_\pi(x_2,b_2)
  \Big] \times \exp[-S_{B}(t_e^1)-S_{\pi}(t_e^1)]
 \nonumber\\ && \;\;+2r_\pi\phi'_\pi(x_2,b_2)\alpha_s(t_e^2)h_e^1(x_2,x_1,b_2,b_1)\times \exp[-S_{B}(t_e^2)-S_{\pi}(t_e^2)]
 \Big\},\label{fe}
  \end{eqnarray}

\item $(V-A)(V+A)$ operators:
  \begin{eqnarray}
  F_{e}^P&=&-32\pi C_Fm_{B}^4r_\pi\int^1_0dx_1dx_2\int^\infty_0b_1db_1b_2db_2
\phi_{B}(x_1,b_1)\cdot \alpha_s(t_e^1)h_e^1(x_1,x_2,b_1,b_2)\nonumber\\
  &&\times\Big\{
   \Big[\phi_\pi(x_2,b_2)+(2+x_2)\phi_\pi(x_2,b_2)+(1-2x_2)r_\pi\phi'_\pi(x_2,b_2)
  \Big]\times \exp[-S_{B}(t_e^1)-S_{\pi}(t_e^1)]
 \nonumber\\ && \;\;+x_1\phi_\pi(x_2,b_2)+2(1-x_1)r_\pi\phi'_\pi(x_2,b_2)\alpha_s(t_e^2)h_e^1(x_2,x_1,b_2,b_1)\times \exp[-S_{B}(t_e^2)-S_{\pi}^1(t_e^2)]
 \Big\},\label{fep}
  \end{eqnarray}

  \begin{eqnarray}
  F_{a}^P&=&-64\pi C_Fm_{B}^4r_\pi\int^1_0dx_1dx_2\int^\infty_0b_1db_1b_2db_2
\phi_{B}(x_1,b_1)\alpha_s(t_a)h_a(x_2,x_3,b_2,b_3)\nonumber\\
  &&\times
   \Big[2\phi_\pi(x_2,b_2)\phi'_\pi(x_3,b_3)+x_2\phi_\pi(x_3,b_3)\phi'_\pi(x_2,b_2)
  \Big] \exp[-S_{\pi}^1(t_a)-S_{\pi}^2(t_a)],\label{fap}
  \end{eqnarray}
 \end{itemize}
where the color factor ${C_F}=4/3$ and $a_i$ represents the corresponding Wilson coefficients from differen decay channels.
$r_\pi=\frac{m_{0}}{m_{B}}=\frac{m_{\pi}^2}{m_{u+m_d}}$. Note that $F_{a}^P$ vanishes in the limit of $m_0=0$. So the $m_0$  term in the pion wave function has a significant impact on the $CP$ violation.

The function are related to the annihilation type process, whose contribution are:
 \begin{eqnarray}
  M_{e}&=&-\frac{32}{3}\pi C_F\sqrt{2N_C}m_{B}^2\int^1_0dx_1dx_2dx_3\int^\infty_0b_1db_1b_2db_2
\phi_{B}(x_1,b_1)\phi_\pi(x_2,b_2)\nonumber\\
  &&\times\phi_\pi(x_3,b_1)\alpha_s(t_d)h_d(x_1,x_2,x_3,b_1,b_2)\exp[-S_{B}(t_d)-S_{\pi}^1(t_d)-S_{\pi}^2(t_d)],\label{me}
  \end{eqnarray}

 \begin{eqnarray}
 M_{a}&=&-\frac{32}{3}\pi C_F\sqrt{2N_C}m_{B}^2\int^1_0dx_1dx_2dx_3\int^\infty_0b_1db_1b_2db_2
\phi_{B}(x_1,b_1)\nonumber\\
 &&\times\Big\{-\Big\{x_2\phi_{\pi}(x_2,b_2)\phi_{\pi}(x_3,b_2)+(x_2+x_3+2)r_{\eta}^2 \phi'_{\pi}(x_2,b_2)\phi'_{\pi}(x_3,b_2)\Big\}\nonumber\\
  &&
\alpha_s(t_{f^1})h_{f^1}(x_1,x_2,x_3,b_1,b_2)\exp[-S_{B}(t_{f^1})-S_{\pi}^1(t_{f^1})-S_{\pi}^2(t_{f^1})]\nonumber\\
  &&
+\Big\{x_2\phi_{\pi}(x_2,b_2)\phi_{\pi}(x_3,b_2)+(2+x_2+x_3)r_{\eta}^2 \phi'_{\pi}(x_2,b_2)\phi'_{\pi}(x_3,b_2)\Big\}\nonumber\\
 &&
\times\alpha_s(t_{f^2})h_{f^2}(x_1,x_2,x_3,b_1,b_2)\exp[-S_{B}(t_{f^2})-S_{\pi}^1(t_{f^2})-S_{\pi}^2(t_{f^2})]
\Big\}.\label{ma}
 \end{eqnarray}

The explicit expressions of individual decay amplitude $F_{e\eta}$ and $F_{e\eta}^{P1,P2}$, $M_{e\eta}$ and $M_{e\eta}^{P1,P2}$, $F_{a\eta}$ and $F_{a\eta}^{P1,P2}$, $M_{a\eta}$ and $M_{a\eta}^{P1,P2}$ can be found
\begin{itemize}
\item $(V-A)(V-A)$ operators:
  \begin{eqnarray}
  F_{e\eta}&=&8\pi
  C_Fm_{B}^4\int^1_0dx_1dx_3\int^\infty_0b_1db_1b_3db_3
\phi_{B}(x_1,b_1)\nonumber\\
  &&\times\Big\{
   \Big[(1+x_3)\phi_\eta(x_3,b_3)+(1-2x_3)r_\eta(\phi_\eta^p(x_3)+\phi_\eta^t(x_3))
  \Big]\alpha_s(t_e^1)h_e(x_1,x_3,b_1,b_3)\exp[-S_{ab}(t_e^1)]
 \nonumber\\ && \;\;+2r_\eta\phi_\eta^p(x_3,b_3)\alpha_s(t_e^2)h_e(x_3,x_1,b_3,b_1) \exp[-S_{ab}(t_e^2)]
 \Big\},\label{ppefll}
  \end{eqnarray}

\item $(V-A)(V+A)$ operators:
\begin{eqnarray}
  F^{P1}_{e\eta}&=&-F_{e\eta},\label{ppeflr}
\end{eqnarray}

\item $(S-P)(S+P)$ operators:
  \begin{eqnarray}
 F^{P2}_{e\eta}&=& 16\pi  C_Fm_{B}^4r_{\eta(')}\int^1_0dx_1dx_3\int^\infty_0b_1db_1b_3db_3
\phi_{B}(x_1,b_1)\nonumber\\
  &&
  \times\Big\{
  \Big[\phi_\eta(x_3,b_3)+r_\eta((2+x_3)\phi_\eta^p(x_3,b_3)-x_3\phi_\eta^t(x_3,b_3))\Big]\alpha_s(t_e^1)
  h_e(x_1,x_3,b_1,b_3)\exp[-S_{ab}(t_e^1)]\nonumber\\
  &&\;\;\;+x_1\phi_\eta(x_3,b_3)-2(x_1-1)r_\eta\phi_\eta^p(x_3,b_3)\alpha_s(t_e^2)h_e(x_3,x_1,b_3,b_1)\exp[-S_{ab}(t_e^2)]\Big\},\label{ppefsp}
  \end{eqnarray}
\end{itemize}
where $r_i=\frac{m_{0i}}{m_{B}}$ and $m_{0i}$ refers to the chiral scale parameter.
\begin{itemize}
\item $(V-A)(V-A)$ operators:
\begin{eqnarray}
M_{e\eta}&=&\frac{16\sqrt{6}}{3}\pi C_Fm_{B}^4/\int^1_0dx_1dx_2dx_3\int^\infty_0b_1db_1b_2db_2
\phi_{B}(x_1,b_1)\phi_{\eta(')}(x_2,b_2)
\nonumber\\
&&\times
\Big\{\Big[2x_3r_\eta\phi_\eta^t(x_3,b_1)-x_3)\phi_{\eta}(x_3,b_1)\Big]
\alpha_s(t_f)h_f(x_1,x_2,x_3,b_1,b_2) \exp[-S_{cd}(t_f)]\Big\},\label{ppenll}
 \end{eqnarray}

\item $(V-A)(V+A)$  and $(S-P)(S+P)$ operators:
\begin{eqnarray}
M^{P1}_{e\eta}=0,\;\;\;M^{P2}_{e\eta}=-M_{e\eta}.
\end{eqnarray}
\end{itemize}

The functions are related with the annihilation type process, whose contributions are:
\begin{itemize}
\item $(V-A)(V-A)$ operators:
\begin{eqnarray}
F_{a\eta}&=&-8\pi C_Fm_{B}^4\int^1_0dx_2dx_3\int^\infty_0b_2db_2b_3db_3
\phi_{B}(x_1,b_1)\nonumber\\
  &&\times\Big\{
   \Big[x_3\phi_\eta(x_3,b_3)\phi_{\eta(')}(x_2,b_2)+2r_\eta r_{\eta(')}((x_3+1)\phi_{\eta}^p(x_3,b_3)\nonumber\\
  &&+(x_3-1)\phi_{\eta}^t(x_3,b_3))\phi_{\eta(')}^p(x_2,b_2)
  \Big]\alpha_s(t_e^3)h_a(x_2,x_3,b_2,b_3)\exp[-S_{gh}(t_e^3)]
 \nonumber\\ && \;\;-\Big[x_2\phi_\eta(x_3,b_3)\phi_{\eta(')}(x_2,b_2)+2r_\eta r_{\eta(')}((x_2+1)\phi_{\eta(')}^p(x_2,b_2)\nonumber\\
  &&+(x_2-1)\phi_{\eta(')}^t(x_2,b_2))\phi_{\eta}^p(x_3,b_3)
  \Big]\alpha_s(t_e^4)h_a(x_3,x_2,b_3,b_2)\exp[-S_{gh}(t_e^4)]
 \Big\},\label{ppafll}
 \end{eqnarray}

 \item
$(V-A)(V+A)$ operators:
\begin{eqnarray}
F_{a\eta}^{P1}=F_{a\eta},\label{ppaflr}
\end{eqnarray}

 \item $(S-P)(S+P)$ operators:
 \begin{eqnarray}
 F_{a\eta}^{P2}&=&-16\pi C_Fm_{B}^4\int^1_0dx_2dx_3\int^\infty_0b_2db_2b_3db_3\phi_{B}(x_1,b_1)\nonumber\\
  &&\times\Big\{
   \Big[x_3r_\eta(\phi_\eta(x_3,b_3)-\phi_\eta^t(x_3,b_3))\phi_{\eta(')}(x_2,b_2)+2 r_{\eta(')}\phi_\eta^p(x_3,b_3)+\phi_{\eta(')}^p(x_2,b_2)
  \Big]\nonumber\\
  &&\times \alpha_s(t_e^3)h_a(x_2,x_3,b_2,b_3) \exp[-S_{gh}(t_e^3)]+\Big[x_2r_{\eta(')}(\phi_{\eta(')}(x_2,b_2)\-\phi_{\eta(')}^t(x_2,b_2))\phi_{\eta}(x_3,b_3)\nonumber\\
  &&+2 r_{\eta}\phi_{\eta(')}(x_2,b_2)\phi_\eta^p(x_3,b_3)
  \Big]\alpha_s(t_e^4)h_a(x_3,x_2,b_3,b_2) \exp[-S_{gh}(t_e^4)]
 \Big\}.\label{ppafsp}
\end{eqnarray}
\end{itemize}

\begin{itemize}
\item $(V-A)(V-A)$ operators:
\begin{eqnarray}
 M_{a\eta}&=&\frac{16\sqrt{6}}{3}\pi C_Fm_{B}^4/\int^1_0dx_1dx_2dx_3\int^\infty_0b_1db_1b_2db_2\phi_{B}(x_1,b_1)
 \times\Big\{-\Big\{x_2\phi_{\eta}(x_3,b_2)\phi_{\eta(')}(x_2,b_2)\nonumber\\
  &&+r_{\eta}r_{\eta(')}[(x_2+x_3+2)\phi_{\eta(')}^p(x_2,b_2)+(x_2-x_3)\phi_{\eta(')}^t(x_2,b_2)]\phi_{\eta}^p(x_3,b_2)
 +r_{\eta}r_{\eta(')}[(x_2-x_3)\phi_{\eta(')}^p(x_3,b_2)\nonumber\\
  &&+(x_2+x_3-2)\phi_{\eta(')}^t(x_2,b_2)]\phi_{\eta}^t(x_3,b_2)\Big\}
\alpha_s(t_f^3)h_f^3(x_1,x_2,x_3,b_1,b_2) \exp[-S_{ef}(t_f^3)]\nonumber\\
  &&
+\Big\{x_3\phi_{\eta}(x_3,b_2)\phi_{\eta(')}(x_2,b_2)+r_{\eta}r_{\eta(')}[(x_2+x_3)\phi_{\eta(')}^p(x_2,b_2)
+(x_3-x_2)\phi_{\eta(')}^t(x_2,b_2)]\phi_{\eta}^p(x_3,b_2)\nonumber\\
 &&
 +r_{\eta}r_{\eta(')}[(x_3-x_2)\phi_{\eta(')}^p(x_2,b_2)+(x_2+x_3)\phi_{\eta(')}^t(x_2,b_2)]\phi_{\eta}^t(x_3,b_2)\Big\}\nonumber\\
 &&
\times\alpha_s(t_f^4)h_f^3(x_1,x_2,x_3,b_1,b_2) \exp[-S_{ef}(t_f^4)]
\Big\},\label{ppanll}
 \end{eqnarray}

 \item $(V-A)(V+A)$ operators:
 \begin{eqnarray}
 M_{a\eta}^{P1}&=&\frac{16\sqrt{6}}{3}\pi C_Fm_{B}^4/\int^1_0dx_1dx_2dx_3\int^\infty_0b_1db_1b_2db_2
\phi_{B}(x_1,b_1)\nonumber\\
 &&\times\Big\{\Big[(x_3-2)r_{\eta}\phi_{\eta(')}(x_2,b_2)(\phi_{\eta}^p(x_3,b_2)+\phi_{\eta}^t(x_3,b_2))-(x_2-2)r_{\eta(')}\phi_{\eta}(x_3,b_2)
 (\phi_{\eta(')}^p(x_2,b_2)\nonumber\\
 &&+\phi_{\eta(')}^t(x_2,b_2))\Big]
\alpha_s(t_f^3)h_f^3(x_1,x_2,x_3,b_1,b_2) \exp[-S_{ef}(t_f^3)]
-\Big[x_3r_{\eta}\phi_{\eta(')}(x_2,b_2)(\phi_{\eta}^p(x_3,b_2)+\phi_{\eta}^t(x_3,b_2))\nonumber\\
 &&
-x_2r_{\eta(')}\phi_{\eta}(x_3,b_2)(\phi_{\eta(')}^p(x_2,b_2)+\phi_{\eta(')}^t(x_2,b_2))\Big]
\alpha_s(t_f^4)h_f^3(x_1,x_2,x_3,b_1,b_2) \exp[-S_{ef}(t_f^4)]
\Big\},\label{ppanlr}
 \end{eqnarray}

 \item $(S-P)(S+P)$ operators:
 \begin{eqnarray}
 M_{a\eta}^{P2}&=&\frac{16\sqrt{6}}{3}\pi C_Fm_{B}^4/\int^1_0dx_1dx_2dx_3\int^\infty_0b_1db_1b_2db_2
\phi_{B}(x_1,b_1)\times\Big\{\Big\{x_3\phi_{\eta}(x_3,b_2)\phi_{\eta(')}(x_2,b_2)\nonumber\\
  &&+r_{\eta} r_{\eta(')}[(x_2+x_3+2)\phi_{\eta(')}^p(x_2,b_2)+(x_3-x_2)\phi_{\eta(')}^t(x_2,b_2)]\phi_{\eta}^p(x_3,b_2)
 +r_{\eta}r_{\eta(')}[(x_3-x_2)\phi_{\eta(')}^p(x_3,b_2)\nonumber\\
  &&+(x_2+x_3-2)\phi_{\eta(')}^t(x_2,b_2)]\phi_{\eta}^t(x_3,b_2)\Big\}
\alpha_s(t_f^3)h_f^3(x_1,x_2,x_3,b_1,b_2) \exp[-S_{ef}(t_f^3)]\nonumber\\
  &&
-\Big\{x_2\phi_{\eta}(x_3,b_2)\phi_{\eta(')}(x_2,b_2)+r_{\eta}r_{\eta(')}[(x_2+x_3)\phi_{\eta(')}^p(x_2,b_2)
+(x_2-x_3)\phi_{\eta(')}^t(x_2,b_2)]\phi_{\eta}^p(x_3,b_2)\nonumber\\
 &&
 +r_{\eta}r_{\eta(')}[(x_2-x_3)\phi_{\eta(')}^p(x_2,b_2)+(x_2+x_3)\phi_{\eta(')}^t(x_2,b_2)]\phi_{\eta}^t(x_3,b_2)\Big\}\nonumber\\
 &&
\times\alpha_s(t_f^4)h_f^3(x_1,x_2,x_3,b_1,b_2) \exp[-S_{ef}(t_f^4)]
\Big\}.
 \label{ppansp}
 \end{eqnarray}
\end{itemize}

If we change the $\pi$ and $\eta(')$ meson, the corresponding expressions of amplitudes for new diagrams will be similar with those as given in Eqs.(\ref{ppefll})-(\ref{ppansp}), since the $\pi$ and $\eta(')$ are all pseudoscalar mesons and have the similar wave functions \cite{Lu2006}. For example, we can find that
\begin{eqnarray}
F_{a\eta(')}= -F_{a\pi}, \;\;\;\;  F_{a\eta(')}^{P1}= F_{a\pi}^{P1}, \;\;\;\;  F_{a\eta(')}^{P2}= F_{a\pi}^{P2}.
\end{eqnarray}



\end{spacing}
\end{document}